\documentclass[10pt,twocolumn]{article}

\usepackage[T1]{fontenc}
\usepackage{lmodern}
\usepackage[letterpaper,top=0.70in,bottom=0.74in,left=0.70in,right=0.70in]{geometry}
\usepackage{microtype}
\usepackage{amsmath,amssymb}
\usepackage{booktabs}
\usepackage{array}
\usepackage{tabularx}
\usepackage{multirow}
\usepackage{enumitem}
\usepackage{natbib}
\usepackage{xcolor}
\usepackage{hyperref}
\usepackage{url}
\usepackage{tikz}
\usepackage{pgfplots}
\usepackage{pgfplotstable}
\usepackage{balance}
\usepackage{placeins}
\usetikzlibrary{positioning,arrows.meta,calc,fit}
\pgfplotsset{compat=1.18}

\definecolor{StudyBlue}{HTML}{245DA8}
\definecolor{StudyTeal}{HTML}{147B70}
\definecolor{StudyGold}{HTML}{A66A16}
\definecolor{StudyRed}{HTML}{B84A3A}
\definecolor{StudyInk}{HTML}{18212B}
\definecolor{StudyMuted}{HTML}{5E6A75}
\definecolor{StudyGrid}{HTML}{D5DCE3}
\definecolor{StudyFill}{HTML}{EFF4F8}

\hypersetup{
  pdftitle={Beyond the Final Prompt: Measuring the Effect of Within-Conversation Context on AI Answers},
  pdfauthor={Benjamin Tannenbaum},
  colorlinks=true,
  linkcolor=StudyBlue,
  citecolor=StudyBlue,
  urlcolor=StudyBlue
}

\setlist{nosep,leftmargin=*}

\newcolumntype{Y}{>{\raggedright\arraybackslash}X}
\newcommand{\pct}[1]{#1\%}

\title{\vspace{-0.45in}\textbf{Beyond the Final Prompt:}\\
Measuring the Effect of Within-Conversation Context on AI Answers}

\author{
  Benjamin Tannenbaum\\
  Aiso, Tel Aviv, Israel\\
  \texttt{ben@getaiso.com}
}

\date{\vspace{-0.12in}}

\begin{document}
\maketitle

\begin{abstract}
An isolated final user message is often treated as the query in evaluations of
AI systems. In a conversation, however, the actionable request may be
distributed across preceding turns. We directly test whether that omitted
within-conversation context changes answers. For each of 180 English
multi-turn conversations sampled from a governed commercial corpus and the
public PRISM dataset, we hold the final user message and requested answer model
constant while generating three answers: one from the full role-labelled
conversation, one from the final message alone, and one from the final message
plus a prefix-only reconstruction capped at 160 words. A separately requested
judge model evaluates answers under randomized labels. The prespecified
primary endpoint is a material difference that could change what the user
does, rather than a difference in style or detail. After inverse-probability
weighting to the eligible cohorts, the full-conversation and isolated-final
answers differ materially in \pct{44.7} of cases (95\% bootstrap CI
\pct{33.8}--\pct{56.1}). Full-conversation answers score 0.49 points higher on
a 0--4 request-satisfaction scale (0.32--0.67). Adding the compressed prefix
reduces the material-difference rate to \pct{30.8}
(\pct{20.2}--\pct{42.1}), a 13.9-point reduction
(\pct{4.9}--\pct{24.1}), and reduces the mean satisfaction gap to 0.01 points
($-0.12$--0.13). Yet compression is not equivalent to the complete dialogue
context: almost one third of answers remain materially different. The primary
comparison is stronger in the commercial cohort (\pct{68.5}) than in PRISM
(\pct{35.4}). An order-swapped repeat on 48 cases yields \pct{91.7} agreement
and $\kappa=0.83$ for the primary decision. The results identify the
conversation, not its endpoint string, as the defensible query unit for many
AI-answer measurements. They concern preceding turns in the same conversation
and do not test persistent memory across separate conversations.
\end{abstract}

\section{Introduction}

What is the query in a conversation with an AI system? A common operational
answer is the latest user message. It is easy to store, replay, count, and place
in a prompt panel. That convenience does not guarantee construct validity. A
user can introduce a goal, receive alternatives, add a budget, reject an
assumption, and end with ``Which one would you choose?'' The final message is
not self-contained. It is a state update whose interpretation depends on the
conversation prefix.

The preceding paper in this sequence measured that distribution directly
\citep{tannenbaum2026prompt}. Across commercial and public conversations, it
found that explicit goals, constraints, alternatives, corrections, and
evidence requirements frequently appeared before the final user message and
were not restated there. That finding established an information-availability
problem. It did not show that the missing context changes what an AI answers.
The present paper closes that empirical gap.

We ask one deliberately narrow question:
\begin{quote}
\emph{Holding the final user message constant, how do the preceding turns in
the same conversation change the answer?}
\end{quote}
The scope is within-conversation context. We do not use ``memory'' to denote
the prefix because that term can imply a persistent store spanning separate
conversations. The available data cannot observe or evaluate such a system.

The experiment is paired at the conversation level. Every case produces an
answer under the full conversation, the isolated final message, and the final
message augmented with a compressed reconstruction of relevant preceding
turns. The reconstruction model never sees the final message, preventing it
from reverse-engineering a convenient summary from the endpoint. The answer
model, decoding interface, and final message are fixed. A separate model judges
all three answers against the observed conversation under randomized labels.

The paper makes four contributions:
\begin{itemize}
  \item a direct paired estimate of how often preceding turns materially
  change an answer when the final message is held constant;
  \item a behavioral definition of material change that excludes wording,
  tone, organization, and detail alone;
  \item a test of compressed request-state reconstruction as a lower-cost
  alternative to carrying the complete conversation prefix; and
  \item a reproducible, privacy-preserving evaluation pipeline with public
  text-free case measures, bootstrap inference, and answer-order reliability.
\end{itemize}

The central result is not that every answer needs every earlier token. It is
that the endpoint string is often an inadequate experimental unit. In the
weighted pooled estimate, final-message isolation materially changes the
answer in \pct{44.7} of cases. Compression closes the average satisfaction gap
but leaves a \pct{30.8} material-difference rate. A concise state summary can
be useful without being interchangeable with the dialogue context that
produced it.

\section{Related Work}

\subsection{Conversational information seeking and rewriting}

Conversational information seeking treats an information need as an
interaction-level object \citep{zamani2023conversational}. Contextual question
rewriting makes a dependent turn usable by a retriever by producing a
self-contained query \citep{elgohary2019unpack}. Later systems generate or
select relevant context to improve reformulation, including ConvGQR and CHIQ
\citep{mo2023convgqr,mo2024chiq}. These methods optimize retrieval inputs. Our
question is downstream and behavioral: if the final message is answered with
or without its conversation prefix, does the answer materially change?

The difference matters because a reconstruction can preserve explicit request
state while losing discourse evidence. Whether an assistant proposal was
accepted, merely acknowledged, corrected later, or still unresolved can
depend on ordering and role labels. Our compressed-prefix condition therefore
tests an intentionally practical approximation rather than assuming that
rewriting is lossless.

\subsection{Context use in language models}

Earlier perturbation studies found that neural dialogue systems did not always
use preceding turns effectively \citep{sankar2019history}. More recent work
shows that language models can degrade when information is disclosed
progressively across turns \citep{laban2025lost}. MT-OSC proposes a more
controlled multi-turn path for such settings \citep{singh2026mtosc}, while
Tokengeist traces the contribution of turns in agentic conversations
\citep{tang2026tokengeist}. \citet{huang2026ownwords} distinguish user-authored
context from a model's own prior words and report that prior assistant text can
sometimes be unnecessary or harmful.

These findings motivate competing predictions. More context can supply
missing constraints and discourse state, but it can also distract the model or
anchor it to obsolete material. Our outcome is therefore symmetric: a
material difference does not automatically mean that the full-conversation
answer is better. Request-satisfaction scores and directional comparisons are
reported separately.

\subsection{The unit of measurement in AI search}

Conversational answers can compress multiple conventional query and source
actions into one response \citep{tannenbaum2026density}. The companion
request-state study shows a complementary input-side compression: a short
message can rely on a specification accumulated over earlier turns
\citep{tannenbaum2026prompt}. The current intervention connects these two
observational claims. If the same endpoint string produces materially
different answers when its prefix changes, answer visibility, brand citation,
factuality, and recommendation measurements cannot generally be attributed to
that string alone.

Model-based evaluation makes paired naturalistic experiments feasible at this
scale, but it introduces its own measurement layer. Work on LLM-as-judge
evaluation documents both strong agreement and systematic biases
\citep{zheng2023judge}. We reduce position effects through deterministic label
randomization and a prespecified order-swapped repeat. This is a reliability
check, not a substitute for independent human validation.

\section{Estimand and Experimental Design}

\subsection{Three answer conditions}

Let a conversation contain role-labelled messages
$m_1,\ldots,m_{T-1},u_T$, where $u_T$ is the final user message. Let $P_T$
denote the complete prefix $m_1,\ldots,m_{T-1}$, including user and assistant
turns. Let $R(P_T)$ be a reconstruction generated from the prefix alone. For a
fixed requested answer model $f$, the three observed outputs are
\begin{align*}
Y_F &= f(P_T,u_T) && \text{full conversation},\\
Y_I &= f(u_T) && \text{isolated final},\\
Y_C &= f(R(P_T),u_T) && \text{compressed prefix}.
\end{align*}

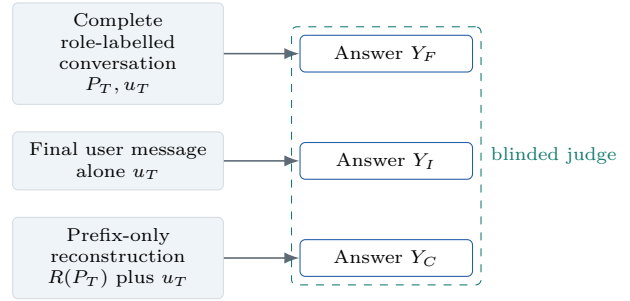
\begin{figure}[t]
\centering
\begin{tikzpicture}[
  font=\scriptsize,
  node distance=3.5mm,
  box/.style={draw=StudyGrid, rounded corners=2pt, fill=StudyFill,
    align=center, inner xsep=5pt, inner ysep=4pt, text width=0.28\columnwidth},
  answerbox/.style={draw=StudyBlue, rounded corners=2pt, fill=white,
    align=center, inner xsep=4pt, inner ysep=4pt, text width=0.23\columnwidth},
  arrow/.style={-{Latex[length=2mm]}, semithick, draw=StudyMuted}
]
\node[box] (full) {Complete role-labelled\\conversation $P_T,u_T$};
\node[box, below=of full] (iso) {Final user message\\alone $u_T$};
\node[box, below=of iso] (comp) {Prefix-only reconstruction\\$R(P_T)$ plus $u_T$};
\node[answerbox, right=10mm of full] (yf) {Answer $Y_F$};
\node[answerbox, right=10mm of iso] (yi) {Answer $Y_I$};
\node[answerbox, right=10mm of comp] (yc) {Answer $Y_C$};
\draw[arrow] (full) -- (yf);
\draw[arrow] (iso) -- (yi);
\draw[arrow] (comp) -- (yc);
\node[draw=StudyTeal, dashed, rounded corners=2pt, fit=(yf)(yi)(yc),
  inner sep=3pt, label={[text=StudyTeal]right:blinded judge}] {};
\end{tikzpicture}
\caption{Paired manipulation. Only the representation of preceding turns
changes. The final user message is held constant.}
\label{fig:design}
\end{figure}

The paired primary contrast is $Y_F$ versus $Y_I$. The compressed contrast
$Y_F$ versus $Y_C$ asks whether a concise reconstruction can reproduce the
answer behavior associated with the full context. Figure~\ref{fig:design}
shows the manipulation. The final user message, requested answer model, output
schema, runner, and answer instructions are identical across conditions.

The runner sends a research instruction followed by tagged quoted data. It
instructs the model not to use tools, to answer directly, to ask a concise
clarifying question rather than guess when essential information is missing,
and to stay under 300 words unless the requested deliverable requires more.
This is a controlled serialization through a model runner, not a replay in a
native consumer chat product.

\subsection{Compressed reconstruction}

The reconstruction instruction asks for the accumulated request context in at
most 160 words. It explicitly names user goals, constraints, alternatives,
corrections, accepted or rejected choices, evidence requirements, and
unresolved questions. It also instructs the model to distinguish assistant
proposals from user requirements and to add no facts. Crucially, reconstruction
uses $P_T$ without access to $u_T$. This design prevents endpoint-conditioned
summarization from copying or rationalizing the final message. Across the 180
cases, valid reconstructions range from 42 to 158 space-separated words, with
a median of 111.

\subsection{Outcomes}

The primary endpoint is a blinded binary judgment that $Y_F$ and $Y_I$
differ materially. A difference is material only if it can change what the
user does through at least one of the following:
\begin{itemize}
  \item a different recommendation, selected item, or factual bottom line;
  \item different compliance with an explicit constraint;
  \item a different requested deliverable; or
  \item different clarification or refusal behavior.
\end{itemize}
Wording, tone, organization, and amount of detail alone are explicitly
excluded. The same rule defines the $Y_F$ versus $Y_C$ secondary endpoint.

The judge also assigns each answer a request-satisfaction score from 0
(unusable or contradictory) to 4 (fully satisfies), plus pairwise severity
from 0 (none) to 3 (outcome-reversing). We report changes in recommendation or
conclusion, clarification or refusal, and constraint compliance. Lexical
distance is one minus Jaccard overlap between lowercased word sets. It is a
descriptive surface measure, not a semantic endpoint.

\subsection{Blinding and reliability}

For each case, a separately requested judge model sees the complete observed
conversation and the three candidate answers under labels A, B, and C. The
mapping from condition to label is deterministically randomized by case. The
judge returns all answer scores and all three pairwise comparisons in a strict
JSON schema. A hash-selected 30\% target subset is judged again under a
different label order; the deterministic selection produces 48 cases
(26.7\%). We calculate percent agreement and Cohen's $\kappa$ for the binary
material-change decisions.

\subsection{Frozen analysis}

The design, conditions, sampling rules, endpoint definition, requested models,
and analysis were frozen before outcome generation. Sampling is stratified by
source, conversation depth, and a transcript-derived dependency heuristic.
Because equal numbers are drawn from unequal eligible strata, the primary
estimand is the inverse-probability-weighted proportion in the pooled eligible
cohorts. Confidence intervals use 5,000 case-bootstrap draws, resampling
within source. We also report unweighted, source-specific, and dependency-band
descriptives. All intervals are percentile intervals and are not adjusted for
multiple secondary comparisons.

\section{Data and Sampling}

\subsection{Corpora}

The commercial source combines a governed discovery export with additional
non-overlapping governed exports. Loader rules first deduplicate identifiers,
remove overlap between discovery and replication files, and remove exact
transcript duplicates. Of 1,477 loaded English-candidate conversations, 544
meet the frozen experimental eligibility rules.

The public source is PRISM, a participatory dataset spanning varied
interaction protocols and participant preferences \citep{kirk2024prism}. The
loader reads 7,125 conversations after released corpus-level exclusions. To
limit participant clustering, the experiment retains at most one eligible
conversation per participant, leaving 1,384 eligible cases.

Eligibility requires at least two user turns, at least one preceding assistant
turn, a final message of 2--96 words, and an approximate full context no longer
than 12,000 tokens. We exclude non-English and unparseable records, pure
acknowledgments or closures, released moderation or personally identifying
information flags, and transcript-detected categories involving contact PII,
self-harm, explicit sexual content, dangerous wrongdoing, or high-stakes
medical, legal, and financial requests. These rules reduce both privacy risk
and the cost of evaluating domains where model judgments would need specialist
review.

\begin{table}[t]
\centering
\small
\begin{tabular}{@{}lrrr@{}}
\toprule
Source & Eligible & Sampled & Weighted share\\
\midrule
Commercial & 544 & 90 & 28.2\%\\
PRISM & 1,384 & 90 & 71.8\%\\
\midrule
Total & 1,928 & 180 & 100.0\%\\
\bottomrule
\end{tabular}
\caption{Eligible cohorts and paired experimental sample. Weighting restores
the pooled eligible-cohort composition after equal source sampling.}
\label{tab:sample}
\end{table}

\subsection{Stratified sample}

The deterministic sample contains 90 commercial and 90 PRISM conversations.
It includes 64 short conversations (2--3 user turns), 62 medium conversations
(4--6 turns), and 54 long conversations (7 or more turns). The dependency
heuristic yields 59 low-control, 63 medium, and 58 high cases. The heuristic
uses only transcript features fixed before answer generation, such as
referential language and constraints present before but not in the final turn.
It supports coverage and descriptive analysis; it is not treated as ground
truth for semantic dependence.

All released case keys are study-specific hashes. Raw transcripts, source
identifiers, model reconstructions, answers, and judge rationales remain in an
ignored private directory. Public per-case artifacts contain only text-free
features, sampling weights, and numerical outcomes. Examples in this paper are
constructed rather than copied from either corpus.

\section{Implementation}

The requested answer and reconstruction model is \texttt{gpt-5.4-mini}; the
requested judge model is \texttt{gpt-5.5}. All calls use the OpenAI Codex CLI
in ephemeral, read-only mode with user configuration and repository rules
ignored. Reasoning effort is set to \texttt{none}. We identify requested model
names rather than claiming a provider-side model snapshot that the interface
does not expose.

Every generation is schema-constrained and checkpointed. The experiment
creates 180 reconstructions, 540 answers, 180 main judgments, and 48 repeat
judgments. Each answer condition receives one stochastic generation per case.
Twelve workers run concurrently, with up to three retries for interface or
schema failure. Prompts wrap all conversation content as quoted data and repeat
an instruction not to follow embedded attempts to redirect the evaluator.

Source provenance records SHA-256 hashes for 46 input files, the frozen JSON
configuration, and the analysis program. Seven unit tests cover label-order
mapping, weighted estimates, deterministic bootstrap behavior, lexical
distance, agreement, exclusion of raw judge files, and the guard preventing
raw-text keys from entering public outputs.

\section{Results}

\subsection{Preceding turns materially change answers}

The primary weighted estimate is \pct{44.7}: nearly half of full-conversation
answers differ materially from answers to the same final user message in
isolation (95\% CI \pct{33.8}--\pct{56.1}). The unweighted paired proportion
is \pct{53.3}. The difference reflects the deliberate equal source sample and
the larger PRISM eligible cohort, not missing cases.

The full-conversation answer scores 0.49 points higher than the isolated-final
answer on the 0--4 satisfaction scale (0.32--0.67). Its weighted mean score is
3.80, compared with 3.31 in isolation. Full context scores higher in
\pct{36.2} of eligible-cohort-weighted cases and lower in \pct{6.0}; the
remaining cases tie. Thus, the symmetric material-change endpoint is
accompanied by a directional quality difference favoring the complete
conversation in this runner.

\begin{table*}[t]
\centering
\small
\begin{tabular}{@{}p{2.65in}rrr@{}}
\toprule
Outcome & Estimate & 95\% CI low & 95\% CI high\\
\midrule
Material change, full vs. isolated & 44.7\% & 33.8\% & 56.1\%\\
Material change, full vs. compressed & 30.8\% & 20.2\% & 42.1\%\\
Rate difference, isolated minus compressed & 13.9 pp & 4.9 pp & 24.1 pp\\
Satisfaction delta, full minus isolated & 0.488 & 0.319 & 0.668\\
Satisfaction delta, full minus compressed & 0.015 & $-0.119$ & 0.133\\
Recommendation or conclusion changed, full vs. isolated & 26.7\% & 17.0\% & 36.9\%\\
Clarification or refusal changed, full vs. isolated & 14.4\% & 8.0\% & 22.0\%\\
Constraint compliance changed, full vs. isolated & 10.6\% & 5.3\% & 17.0\%\\
\bottomrule
\end{tabular}
\caption{Inverse-probability-weighted estimates for the pooled eligible
cohorts. Intervals use 5,000 within-source case-bootstrap draws.}
\label{tab:primary}
\end{table*}

The judge attributes \pct{26.7} of primary-pair differences to a changed
recommendation or conclusion, \pct{14.4} to changed clarification or refusal
behavior, and \pct{10.6} to changed constraint compliance. Categories can
co-occur. Mean pairwise severity is 0.65 on the 0--3 scale. Weighted answer
length rises from 96 words in isolation to 110 words with the full
conversation, but length alone cannot explain a primary decision because the
rubric excludes detail-only differences.

\subsection{Compression helps but is not equivalent}

Adding the prefix-only reconstruction lowers the material-difference rate from
\pct{44.7} to \pct{30.8}. The paired rate reduction is 13.9 percentage points
(\pct{4.9}--\pct{24.1}), or 31.1\% relative to the isolated rate. The
unweighted rates are \pct{53.3} and \pct{33.3}, a 20.0-point difference.

Compression has a larger effect on satisfaction than on answer identity. The
compressed answer has a weighted mean satisfaction of 3.78, compared with
3.80 under the full conversation. The mean difference is 0.01 points and its
interval includes zero. Full context scores higher than compression in
\pct{11.4} of weighted cases and lower in \pct{8.5}. Despite this parity,
\pct{30.8} of answer pairs remain materially different. Equivalent average
quality therefore does not imply equivalent recommendations, conclusions, or
interaction behavior.

Surface similarity tells the same limited story. Mean lexical distance from
the full answer is 0.707 for isolation and 0.650 for compression. The smaller
distance is consistent with partial recovery, but both values are large and
lexical overlap cannot determine whether two answers lead to the same action.
Compressed answers average 114 words, slightly longer than full-context
answers.

\subsection{Source and dependency variation}

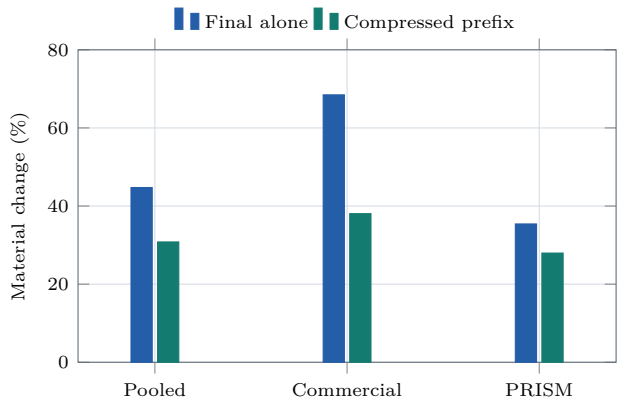
\begin{figure}[t]
\centering
\begin{tikzpicture}
\begin{axis}[
  ybar,
  width=\columnwidth,
  height=2.25in,
  ymin=0,ymax=80,
  ylabel={Material change (\%)},
  symbolic x coords={Pooled,Commercial,PRISM},
  xtick=data,
  xticklabel style={font=\scriptsize},
  yticklabel style={font=\scriptsize},
  ylabel style={font=\scriptsize},
  bar width=8pt,
  enlarge x limits=0.20,
  legend style={font=\scriptsize,draw=none,fill=none,at={(0.5,1.02)},anchor=south,legend columns=2},
  grid=major,
  grid style={draw=StudyGrid},
  axis line style={draw=StudyMuted},
]
\addplot[fill=StudyBlue,draw=StudyBlue] coordinates {
  (Pooled,44.744) (Commercial,68.474) (PRISM,35.417)};
\addplot[fill=StudyTeal,draw=StudyTeal] coordinates {
  (Pooled,30.809) (Commercial,38.051) (PRISM,27.962)};
\legend{Final alone,Compressed prefix}
\end{axis}
\end{tikzpicture}
\caption{Weighted material-change rates relative to the full-conversation
answer. Source differences are descriptive, not randomized.}
\label{fig:source}
\end{figure}

Figure~\ref{fig:source} separates sources. In the commercial cohort, the
full-versus-isolated material-change rate is \pct{68.5}, compared with
\pct{35.4} in PRISM. Compression reduces the commercial rate to \pct{38.1}, a
30.4-point reduction, but reduces the PRISM rate only to \pct{28.0}, a
7.5-point reduction. Full-context satisfaction exceeds isolated-final
satisfaction by 0.91 points in the commercial cohort and 0.32 in PRISM.

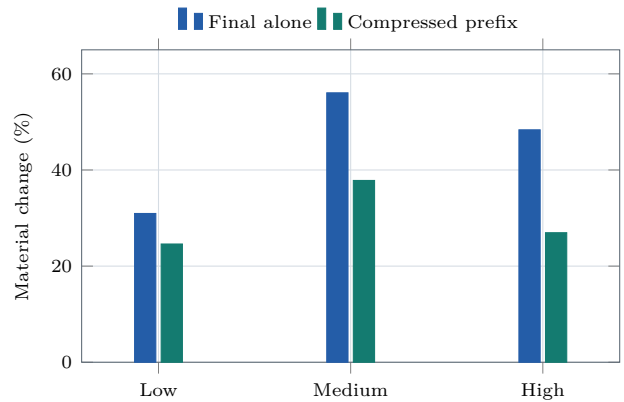
\begin{figure}[t]
\centering
\begin{tikzpicture}
\begin{axis}[
  ybar,
  width=\columnwidth,
  height=2.25in,
  ymin=0,ymax=65,
  ylabel={Material change (\%)},
  symbolic x coords={Low,Medium,High},
  xtick=data,
  xticklabel style={font=\scriptsize},
  yticklabel style={font=\scriptsize},
  ylabel style={font=\scriptsize},
  bar width=8pt,
  enlarge x limits=0.20,
  legend style={font=\scriptsize,draw=none,fill=none,at={(0.5,1.02)},anchor=south,legend columns=2},
  grid=major,
  grid style={draw=StudyGrid},
  axis line style={draw=StudyMuted},
]
\addplot[fill=StudyBlue,draw=StudyBlue] coordinates {
  (Low,30.952) (Medium,56.054) (High,48.356)};
\addplot[fill=StudyTeal,draw=StudyTeal] coordinates {
  (Low,24.603) (Medium,37.823) (High,26.951)};
\legend{Final alone,Compressed prefix}
\end{axis}
\end{tikzpicture}
\caption{Weighted rates by frozen context-dependency band. Sample sizes are
59 low-control, 63 medium, and 58 high.}
\label{fig:dependency}
\end{figure}

The result is also present across all three transcript-derived dependency
bands (Figure~\ref{fig:dependency}). Final-message isolation changes
\pct{31.0} of low-control answers, \pct{56.1} of medium-band answers, and
\pct{48.4} of high-band answers. Compression lowers these rates to
\pct{24.6}, \pct{37.8}, and \pct{27.0}, respectively. The non-monotonic
isolated rates warn against treating a transparent lexical heuristic as an
oracle for behavioral context dependence. Even the low-control band is not a
safe set of self-contained final messages.

\subsection{Judge order reliability}

On the 48-case order-swapped repeat, the primary material-change decision has
\pct{91.7} agreement and Cohen's $\kappa=0.83$. The full-versus-compressed
decision has \pct{85.4} agreement and $\kappa=0.70$. These values indicate
stronger stability for the primary comparison than for the closer compressed
comparison. They do not measure agreement with human experts, and repeated
judgments are not independent because they use the same requested judge model
and rubric.

\section{Interpretation}

\subsection{The conversation is the query unit}

Holding the endpoint constant isolates a basic measurement error. If an
evaluation replays $u_T$ without $P_T$, it is not merely shortening the input;
in \pct{44.7} of eligible-cohort-weighted cases it elicits a materially
different answer. For a visibility tracker, that can mean a different cited
brand. For a recommendation benchmark, it can mean a different selected item.
For a support evaluation, it can mean answering instead of clarifying. The
measured object is a property of the conversation-conditioned request, not of
the final string alone.

This does not imply that every preceding turn should be passed to every model
indefinitely. It implies that an evaluation must define and preserve the
request context that makes the final message interpretable. If a production
system uses truncation, summarization, retrieval over preceding turns, or a
native dialogue state, the evaluation should reproduce that context policy.
Otherwise it estimates behavior for a different system.

\subsection{Compression recovers quality more than identity}

The compressed-prefix condition reveals a useful asymmetry. It reduces the
average satisfaction gap from 0.49 to 0.01 points, yet almost one third of
answers remain materially different from the full-context answer. A
reconstruction can capture enough explicit constraints to support a good
answer while still changing which reasonable answer the model selects.

Several mechanisms are consistent with this pattern. Compression can discard
the order in which alternatives were proposed and rejected. It can turn
assistant suggestions into apparently settled facts despite an instruction to
preserve roles. It can omit hesitation, unresolved ambiguity, or evidence that
shapes emphasis. It can also remove distracting or obsolete content, which
may explain the small set of cases where compressed or isolated answers score
higher. The experiment identifies the net behavioral difference but does not
attribute individual cases to these mechanisms.

\subsection{Implications for evaluation design}

Three practical consequences follow. First, prompt sets derived from final
messages should be labelled as endpoint-message evaluations, not conversation
evaluations. Second, conversational benchmarks should store role-labelled
prefixes or a versioned context-construction policy. Third, metrics should
separate answer quality from answer identity. A compression method can retain
mean judged satisfaction while altering recommendations in ways that matter
for product, policy, or commercial measurement.

The source difference also cautions against transferring a single benchmark
rate. The governed commercial cohort has nearly twice the isolated-final
material-change rate of PRISM. This may reflect task composition, conversation
structure, selection, or model-user interaction patterns. Source is not
randomized, so the difference is descriptive. The robust claim is narrower:
material effects appear in both sources and the pooled estimate remains large
after weighting.

\section{Limitations}

\paragraph{One answer model and one runner.}
The intervention requests one answer model through the OpenAI Codex CLI. Model
families, snapshots, native chat applications, system instructions, tool use,
and context-management policies can yield different effects. The full
condition is a role-labelled serialization inside a research instruction, not
a native message-array API replay.

\paragraph{One generation per condition.}
Each case-condition pair has one stochastic output. Pairing controls the input
comparison but does not estimate within-condition generation variance. A
hierarchical design with repeated outputs would separate context effects from
sampling variance and would likely widen some intervals.

\paragraph{Model-based judgment.}
A different requested model evaluates the answers, but it remains an automated
judge. Label randomization and order-swap reliability address position
sensitivity, not shared model biases or domain expertise. Human adjudication
on a privacy-cleared sample is an important replication.

\paragraph{Compression is one operationalization.}
The 160-word prefix-only reconstruction is intentionally constrained and uses
the same requested model family as answer generation. Longer summaries,
extractive state stores, human-authored rewrites, or retrieval over specific
turns may perform differently. The result does not establish an optimal
compression method.

\paragraph{Sampling and weighting.}
The commercial corpus is governed and naturalistic but not publicly
redistributable. PRISM is public but collected under specific protocols. Equal
source sampling improves comparison, while inverse-probability weighting makes
the pooled estimate depend more heavily on the larger PRISM eligible cohort.
The confidence intervals capture case resampling under the frozen design, not
uncertainty about corpus construction or model version.

\paragraph{Scope of context.}
The study concerns preceding turns inside the same observed conversation. It
does not evaluate persistent memory, account-level personalization, retrieved
records from other sessions, or hidden application state. Those mechanisms
require different data and interventions.

\paragraph{Safety exclusions.}
High-stakes and sensitive categories are excluded. This improves privacy and
reduces inappropriate automated evaluation, but it means the estimates should
not be extended to medical, legal, financial, self-harm, sexual, dangerous, or
contact-PII interactions.

\section{Ethics, Privacy, and Reproducibility}

The experiment sends sampled conversation prefixes to the named model
interface only after explicit authorization for this study. Raw text and model
outputs are never placed in public artifacts or the arXiv package. Public
results contain study-hashed case keys, source-level labels, coarse bands,
weights, and numerical measures. The commercial source cannot be redistributed;
PRISM remains governed by its original release terms.

The accompanying project archive includes the frozen design, configuration,
schemas, experiment and analysis code, unit tests, source-file hashes,
aggregate results, and text-free per-case metrics. These materials support
auditing of the sample, estimand, weighting, bootstrap, and figures without
exposing conversations. Reproduction of model outputs additionally requires
authorized access to the governed inputs and the requested model interfaces.

\section{Conclusion}

The final user message is not generally the whole query. With that message
held constant, removing preceding turns materially changes \pct{44.7} of
answers in the weighted eligible cohorts and lowers judged request satisfaction
by nearly half a point on a 0--4 scale. A compressed reconstruction recovers
most of the average satisfaction gap and reduces the material-change rate by
13.9 points, but it remains behaviorally different from the full conversation
in \pct{30.8} of cases.

The appropriate lesson is methodological. Evaluations of conversational AI
must specify the conversation prefix, dialogue state, or accumulated request
context that conditions an answer. Replaying endpoint strings alone measures a
different task. Persistent memory across separate conversations remains a
separate research question.

\appendix
\section{Constructed Illustration}

Consider a user who first asks for a travel laptop, later adds a budget and
Linux-support constraint, rejects one proposed model, and ends with ``Which is
better?'' The isolated final message does not identify the alternatives,
budget, or rejected option. A compressed prefix may restore those explicit
facts, while the complete conversation also preserves which assistant proposed
each option and how the user reacted. This example is constructed and is not a
dataset excerpt.

\section{Public Artifact Boundary}

The public per-case table contains no transcript or generated-answer text. Its
fields cover study-hashed case key, source, depth and dependency bands,
inverse-probability weight, binary judge outcomes, satisfaction scores,
severity, lexical distance, word counts, and counts of retained
constraint-category terms. Judge rationales, raw model JSON, original
identifiers, participant identifiers, and collection filenames remain
private.

\balance
\bibliographystyle{plainnat}
\bibliography{references}

\end{document}